\documentclass[12pt,onecolumn]{IEEEtran}
\usepackage[T1]{fontenc}
\usepackage{amssymb}
\usepackage{amsmath}
\usepackage{amsthm}
\usepackage{amsfonts}
\usepackage{amsthm}
\usepackage{amsmath}
\usepackage{amsfonts}
\usepackage{amssymb}
\usepackage{graphicx}
\usepackage{hyperref}
\usepackage{mathtools}
\usepackage{dsfont}
\usepackage{verbatim}
\usepackage{subcaption}
\usepackage{float}
\usepackage{eucal}
\usepackage{balance}
\usepackage{bm}
\usepackage{amssymb}
\usepackage{amsmath}
\usepackage{amsthm}
\usepackage{amsfonts}
\usepackage{hyperref}
\usepackage{bbold}
\usepackage{mathtools}
\usepackage{dsfont}
\usepackage{verbatim}
\usepackage{subcaption}

\newtheorem{theorem}{Theorem}

\newtheorem{remark}[theorem]{Remark}

\newcommand\blfootnote[1]{%
  \begingroup
  \renewcommand\thefootnote{}\footnote{#1}%
  \addtocounter{footnote}{-1}%
  \endgroup
}

\title{Comments and Corrections to ``Capacity of Multiple-Antenna Systems With Both Receiver and Transmitter Channel State Information''}

\author{
\IEEEauthorblockN{Kamal Singh, Chandradeep Singh}
\vspace*{-0.65cm}
}

\begin{document}
\maketitle
\begin{abstract}
The correspondence cited in the title~\cite{jayaweera} derived ergodic capacity of the coherent multiple-input multiple-output (MIMO) channel in independent and identically distributed (IID) Rayleigh fading. While the theoretical results are correct, several plots in the paper are incorrect. In this correspondence, we correct the plots. More importantly, the corrected plots present an interesting and compelling contrast between performances of the coherent MIMO systems with and without channel state information at the transmitter; whereas this view is somewhat limited in~\cite{jayaweera} because of flaws in the capacity curves.
\end{abstract}
\begin{IEEEkeywords}
Ergodic capacity, multiple-input multiple-output (MIMO), Rayleigh fading channel, power control, channel state information at the transmitter~(CSIT).
\end{IEEEkeywords}

\section{Introduction}
\IEEEPARstart{T}{he} main objective of this note is to correct the ergodic capacity versus SNR graphs of the MIMO fading channel in~\cite{jayaweera}. The ergodic capacity of a coherent MIMO fading channel $\bm{H}$\footnote{The matrix $\bm{H}$ is a $N_R \times N_T$ matrix, where $N_R$ and $N_T$ indicate the number of receive and transmit antennas, respectively.} with perfect instantaneous channel state information $H$ at the transmitter (CSIT) can be obtained by solving the following problem\footnote{Covariance matrix of the receiver AWGN noise vector, generally denoted as $N_0 \bm{I}_{N_R}$ in the literature, is taken to be the identity i.e. $N_0 = 1$.}~\cite[Section~8.2]{tsebook}:
\begin{align}\label{eq:gen:cap}
\max_{Q(H):\text{Tr}(\mathbb{E}_{\bm{H}}[Q(\bm{H})]) \leq P} \,\,\mathbb{E}_{\bm{H}} [\log \det (\bm{I}_{N_R} + \bm{H} Q(\bm{H})\bm{H}^{\dagger})]
\end{align}
where $Q(H)$ is the input covariance matrix and $P$ is an average power constraint at the transmitter, which implies $\text{Tr}(\mathbb{E}_{\bm{H}} [Q(\bm{H})]) \leq P$. In~\cite{jayaweera}, the authors solved the optimization in~\eqref{eq:gen:cap} for the IID Rayleigh fading model, i.e. channel matrix $\bm{H}$ has i.i.d. entries and each entry in $\bm{H}\sim\mathcal{CN}(0,1)$), to yield the capacity
with CSIT as
\begin{align}\label{eq:cap:csit0}
C &= m \, \mathbb{E}_{\boldsymbol{\lambda}} \bigl[\log(1 +\boldsymbol{\lambda} P(\boldsymbol{\lambda}))\bigr]
\end{align}
where $m = \min(N_R,N_T)$ and the optimal \emph{waterfilling} power scheme $P(\lambda) = (1/\lambda_0 - 1/\lambda)^+$.
We will use explicit notation $C(x,y,z)$ (resp. $\widehat{C}(x,y,z)$) to denote the capacity of this $x \times y$ MIMO fading channel with CSIT (resp. without CSIT) under constraint $z$ on the average transmit power. Precisely, the ergodic capacity has an integral-form expression as given in the Eq.~(58) in~\cite{jayaweera}, and is reproduced here with a simple change of variables\footnote{The $\gamma$ variable in the Eq.~(58) in~\cite{jayaweera} and $\lambda$ variable in Eq.~\eqref{eq:cap} above are related as $\gamma = \lambda \tfrac{P}{mN_0}$.} as follows:
\begin{align}\label{eq:cap}
C(N_R,N_T,P) = m \int_{\lambda_0}^{\infty} \log \left(\dfrac{\lambda}{\lambda_0}\right) f_{\boldsymbol{\lambda}} (\lambda) d \lambda
\end{align}
where $\lambda_0$ is the threshold parameter determined from
\begin{align}\label{eq:cutoff}
\int_{\lambda_0}^{\infty}\left( \dfrac{1}{\lambda_0} -\dfrac{1}{\lambda}\right) f_{\boldsymbol{\lambda}}(\lambda) d \lambda = \dfrac{P}{m}
\end{align}
\blfootnote{Software used to generate the numerical results and Monte Carlo simulation results in this paper can be downloaded from GitHub: \url{
https://github.com/kamalsinghsnu/CapacityOfRayleighFadingMIMOchannel}}
and the eigenmode distribution $f_{\boldsymbol{\lambda}} (\lambda)$ is given by
\vspace*{7pt}
\begin{align}\label{eq:eigendist}
f_{\boldsymbol{\lambda}} (\lambda) = \dfrac{e^{-\lambda} \lambda^{n-m} }{m} \sum_{k=0}^{m-1} \dfrac{k!}{(k+n-m)!} [L_{k}^{n-m}(\lambda)]^2\\[-13pt] \notag
\end{align}
where, in turn, $n = \max(N_R,\,N_T)$ and $L_k^{n-m}(\lambda)$, the associated Laguerre polynomial of order $k$, has a closed-form expression given as\vspace*{7pt}
\begin{align}\label{eq:associatedLaguarre:simplification}
L_{k}^{n-m}(\lambda) = \sum_{p = 0}^{k} \dfrac{(-\lambda)^{p}}{p!} {k + n - m \choose k - p} \cdot \\[-13pt] \notag
\end{align}
The ergodic capacity of this coherent MIMO channel \emph{without} CSIT is given by\vspace*{7pt}
\begin{align}\label{eq:cap:nocsit}
\widehat{C}(N_R,N_T,P)= m \int_{0}^{\infty} \log \left(1 + \lambda P/N_T \right) f_{\boldsymbol{\lambda}} (\lambda) d \lambda \\[-13pt] \notag
\end{align}
and the optimal power allocation is obtained by dividing the total transmit power $P$ equally among all transmit antennas~\cite[Theorem~2]{telatar_mimo}. It is easy to verify that\vspace*{7pt}
\begin{align}\label{eq:cap_rev:csit}
  C(N_R,N_T,P) &= C(N_T,N_R,P),\\[-13pt] \notag
\end{align}
and \vspace*{7pt}
\begin{align}\label{eq:cap_rev:no csit1}
\widehat{C}(N_R,N_T,P) \neq \widehat{C}(N_T,N_R,P)\\[-13pt] \notag
\end{align}
holds in general, except with equality when $N_R = N_T$. In fact, it is easy to check that for $N_R \,\,>\,\, N_T$,
\vspace*{7pt}
\begin{align}\label{eq:cap_rev:no csit2}
  \widehat{C}(N_R,N_T,P) &> \widehat{C}(N_T,N_R,P)\,\,\,\text{holds.}\\[-13pt] \notag
  \end{align}
In the following, for simplicity of notation, we will drop the functional dependence so that $C$ and $\widehat{C}$ should be understood to refer to $C(N_R,N_T,P)$ and $\widehat{C}(N_R,N_T,P)$ respectively. In the next section, we present counter ergodic capacity results with detailed justifications followed by comparison with the existing curves in the correspondence cited in the title. Corrections are also applied to the plots for the cut-off and
the outage probability bounds in~\cite{jayaweera}. The note concludes with a brief discussion and implication of the corrected capacity plots with and without CSIT.

\section{Counter results}
\begin{figure}[H]
\centering
\includegraphics[scale = 1.1]{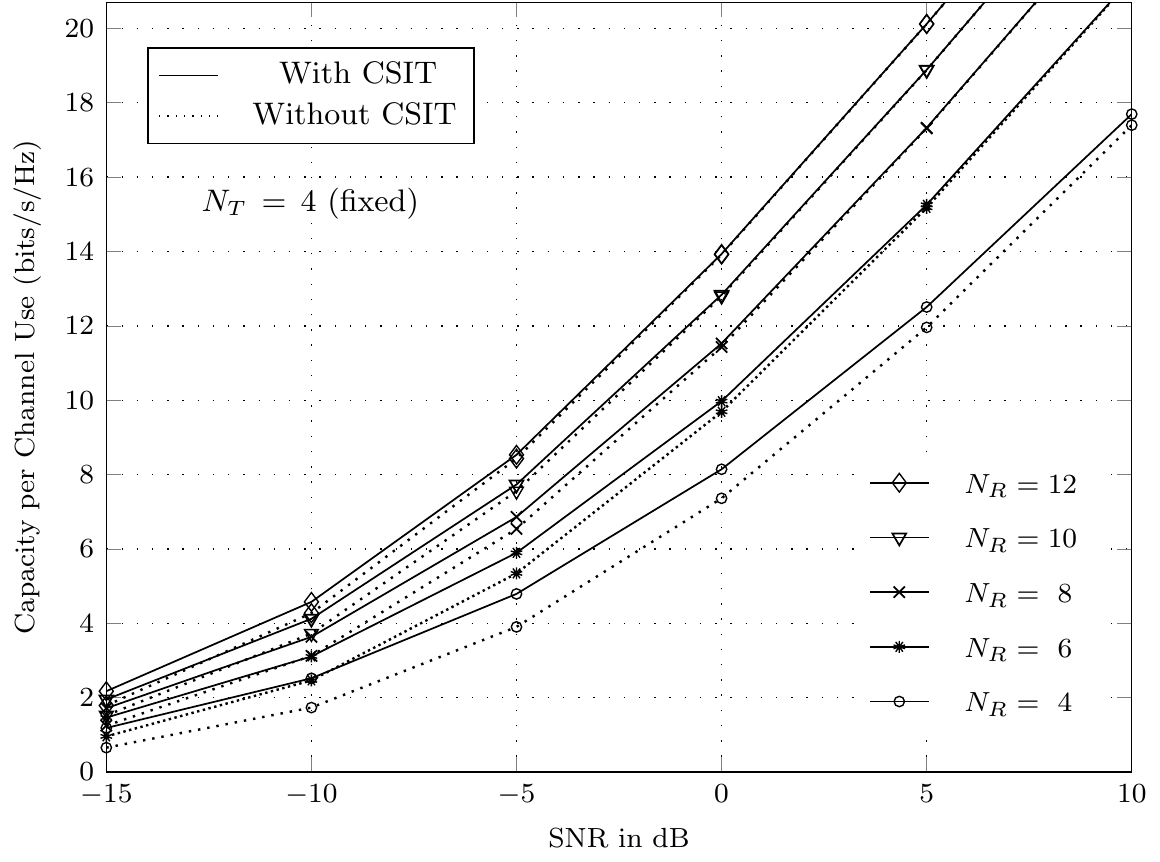}
\caption{Ergodic capacity of \emph{coherent} $N_R \times N_T$ MIMO IID Rayleigh channel with $N_T = 4$ (fixed).}
\label{fig:one}
\end{figure}

\begin{figure}[H]
\centering
\includegraphics[scale = 1.1]{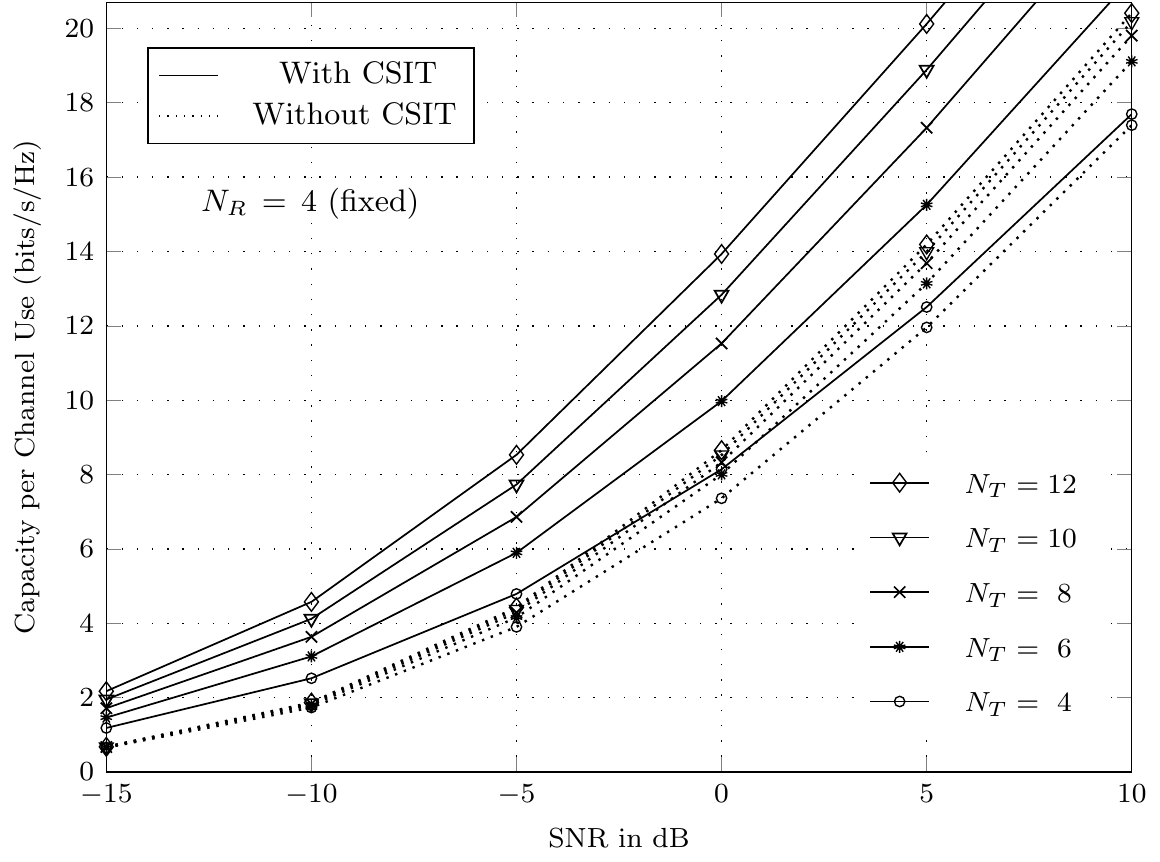}
\caption{Ergodic capacity of \emph{coherent} $N_R \times N_T$ MIMO IID Rayleigh channel with $N_R = 4$ (fixed).}
\label{fig:two}
\end{figure}

The focus is on the ergodic capacity versus SNR curves for the MIMO Rayleigh channel with CSIT presented in the Fig.~5 (or Fig.~6) in~\cite{jayaweera}. In this section, we present counter results for the ergodic capacity of the MIMO Rayleigh channel with CSIT computed using two independent approaches: \emph{one} set of results is computed using standard root-finding algorithms to
evaluate the waterfilling level $1/\lambda_0$ in Eq.~\eqref{eq:cutoff} and then using numerical integration in Eq.~\eqref{eq:cap}, and the \emph{second} set of results using Monte Carlo simulation\footnote{$10^6$ samples are considered for
Monte Carlo simulations.}. The range of SNR\footnote{SNR is taken as $P/(mN_0)$, similar to as Eq.~(34) in~\cite{jayaweera}.}, $N_T$ and $N_R$ are chosen as identical to that in~\cite{jayaweera}. The computed values with these two methods are close and thus, suggesting the correctness of the results; precisely, the values match exactly up to 2 decimal places, see Table~\ref{tab:fading1}-Table~\ref{tab:fading5}.
\begin{table}[H]
\caption{$N_R = 4$, $N_T = 4$ }\label{tab:fading1}
\centering
\begin{tabular}{c | c | c}
\hline 
SNR  &  Capacity (Monte Carlo) &  Capacity (Numerical) \\
(dB)  &(bits/s/Hz) & (bits/s/Hz) \\[.50ex]
\hline
-15&1.182976&1.182924\\\hline
-10&2.518979&2.519749\\\hline
-5&4.788550&4.789042\\\hline
0&8.141317&8.141885\\\hline
5&12.502087&12.506322\\\hline
10&17.699588&17.695610\\\hline
\end{tabular}
\end{table}
\begin{table}[H]
\caption{$N_R = 4$, $N_T = 6$ }\label{tab:fading2}
\centering
\begin{tabular}{c | c | c}
\hline 
SNR  &  Capacity (Monte Carlo) &  Capacity (Numerical) \\
(dB)  &(bits/s/Hz) & (bits/s/Hz) \\[.50ex]
\hline
-15&1.461202&1.461242\\\hline
-10&3.108600&3.108137\\\hline
-5&5.894179&5.894378\\\hline
0&9.973844&9.974640\\\hline
5&15.256107&15.255904\\\hline
10&21.346760&21.346403\\\hline
\end{tabular}
\end{table}
\begin{table}[H]
\caption{$N_R = 4$, $N_T = 8$ }\label{tab:fading3}
\centering
\begin{tabular}{c | c | c}
\hline 
SNR  &  Capacity (Monte Carlo) &  Capacity (Numerical) \\
(dB)  &(bits/s/Hz) & (bits/s/Hz) \\[.50ex]
\hline
-15&1.714074&1.714015\\\hline
-10&3.633365&3.634128\\\hline
-5&6.858986&6.859198\\\hline
0&11.528260&11.527121\\\hline
5&17.324821&17.326976\\\hline
10&23.673975&23.674296\\\hline
\end{tabular}
\end{table}
\begin{table}[H]
\caption{$N_R = 4$, $N_T = 10$ }\label{tab:fading4}
\centering
\begin{tabular}{c | c | c}
\hline 
SNR  &  Capacity (Monte Carlo) &  Capacity (Numerical) \\
(dB)  &(bits/s/Hz) & (bits/s/Hz) \\[0.50ex]
\hline
-15&1.949946&1.949609\\\hline
-10&4.120259&4.120068\\\hline
-5&7.730976&7.732018\\\hline
0&12.835914&12.836162\\\hline
5&18.886564&18.886957\\\hline
10&25.328800&25.329797\\\hline
\end{tabular}
\end{table}
\begin{table}[H]
\caption{$N_R = 4$, $N_T = 12$ }\label{tab:fading5}
\centering
\begin{tabular}{c | c | c}
\hline 
SNR  &  Capacity (Monte Carlo) &  Capacity (Numerical) \\
(dB)  &(bits/s/Hz) & (bits/s/Hz) \\[.50ex]
\hline
-15&2.171655&2.172626\\\hline
-10&4.576217&4.574792\\\hline
-5&8.530615&8.530668\\\hline
0&13.934214&13.933461\\\hline
5&20.122317&20.121336\\\hline
10&26.612371&26.613163\\\hline
\end{tabular}
\end{table}
The capacity values with CSIT in Table~\ref{tab:fading1}-Table~\ref{tab:fading5} are plotted in Fig.~\ref{fig:one} and
Fig.~\ref{fig:two} along with the corresponding capacity values \emph{without} CSIT. The capacity results without CSIT are determined by solving the integral in~\eqref{eq:cap:nocsit} numerically and are also verified with Monte Carlo simulations, see Table~\ref{tab:fading6}-Table~\ref{tab:fading14} in Appendix~\ref{sec:app:A}.
\noindent

To further validate the capacity results with CSIT as shown in Fig.~\ref{fig:one} (or Fig.~\ref{fig:two}), we focus on the extreme SNR regimes.
\begin{itemize}
\item In the \emph{high} SNR regime, the optimal waterfilling scheme allocates nearly same power for \emph{all} states i.e. $P(\lambda) \approx P/m$~\cite[pp.~346-348]{tsebook}.
Thus, at high SNR,
\begin{align}\label{eq:cap:csit1}
C &= m \, \mathbb{E}_{\boldsymbol{\lambda}} \bigl[\log(1 + \boldsymbol{\lambda} P(\boldsymbol{\lambda}))\bigr] \notag \\
                &\approx m \int_{0}^{\infty} \log \left(1 +  \lambda P/m  \right) f_{\boldsymbol{\lambda}} (\lambda) d \lambda.
\end{align}
Comparing~\eqref{eq:cap:csit1} with~\eqref{eq:cap:nocsit}, it is straightforward that, at high SNR,
\begin{align}\label{eq:cap:approxi}
C \approx \widehat{C}
\end{align}
whenever $m = N_T$. This fact, clearly visible in Fig.~\ref{fig:one} at high SNRs, validates the correctness of the capacity values with CSIT presented in this note.
\item In the \emph{low} SNR regime,~\cite{tall} showed that the ergodic capacity with CSIT of this MIMO fading channel scales asymptotically as SNR$\,\log$($1/$SNR)\footnote{$P/N_0$ is taken as SNR in~\cite{tall}.}; in particular, an on-off transmission scheme on the strongest eigenmode\footnote{The CDF of $\boldsymbol{\lambda_{max}}$ is borrowed from~\cite[Corollary 2]{kang}. Alternatively, the PDF of $\boldsymbol{\lambda_{max}}$ is available in~\cite[Eq.~(38)]{zanella}.} (say $\boldsymbol{\lambda_{max}}$) is proposed that is ``asymptotically (at low SNR) capacity-achieving''. This observation is also hinted in~\cite[Eq.~(7.15)]{tsebook}.
    The ergodic rate achievable with the on-off transmission scheme is given by~\cite{tall}
    \begin{align}\label{eq:onoff}
     R &= \mathbb{E}_{\boldsymbol{\lambda_{max}}} \bigl[\log(1 + \lambda_{max} P(\lambda_{max}))\bigr] \notag \\
       &= \int_{\tau}^{\infty} \log \left(1 +  \lambda P_0  \right) f_{\boldsymbol{\lambda_{max}}} (\lambda) d \lambda
    \end{align}
    where $\tau$ is the threshold (chosen same as the \emph{waterfilling} threshold $\lambda_0$) and $P_0 = P/\int_{\tau}^{\infty} f_{\boldsymbol{\lambda_{max}}} (\lambda) d \lambda$. These rates are obtained numerically for a wide range of low SNR values (see Table~\ref{tab:fading15}-Table~\ref{tab:fading19} in Appendix~\ref{sec:app:B}) and plotted in Fig.~\ref{fig:three} along with corresponding ergodic capacity with CSIT achieved using the `optimal' waterfilling scheme over all the non-zero eigenvalues.
    \begin{figure}[H]
\centering
\includegraphics[scale = 1.1]{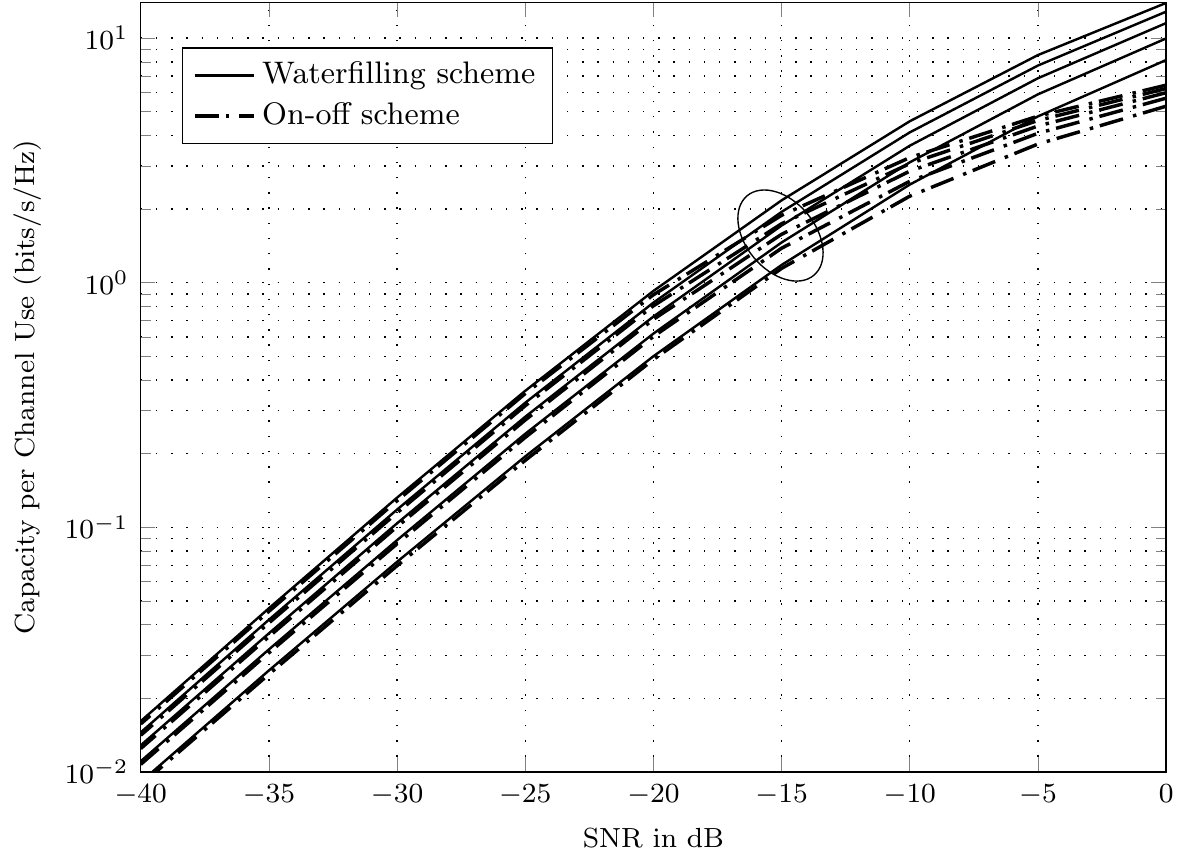}
\caption{Ergodic rates of \emph{coherent} $N_R \times N_T$ MIMO IID Rayleigh channel with CSIT and $N_R = 4$ (fixed) at \emph{low} SNRs. For each power scheme (On-off or Waterfilling), the curves correspond, in ascending order, to $N_T = 4,6,8,10,12$.}
\label{fig:three}
\end{figure}
    The ergodic rates achievable using on-off power scheme are close to the computed capacity values with CSIT at sufficiently low SNRs; for example, observe the rates in Fig.~\ref{fig:three} at the SNR of $-15$ dB or lower~(for exact comparison, see Table~\ref{tab:fading15}-Table~\ref{tab:fading19} in Appendix~\ref{sec:app:B}). This further validates the correctness of the capacity results with CSIT presented here.
\end{itemize}

A close examination of the ergodic capacity versus SNR curves in the Fig.~5 and Fig.~6 in~\cite{jayaweera},
when compared with Fig.~\ref{fig:one} and Fig.~\ref{fig:two} respectively in this note, indicates several differences as follows:
\begin{itemize}
\item At low SNRs, the capacity values with CSIT in the~Fig.~5 in~\cite{jayaweera} show significantly `larger' improvements with increasing $N_R$ receive antennas than the corresponding values in~Fig.~\ref{fig:one} in this note. For example, at low enough SNR of $-15$ dB, the capacity with CSIT in the~Fig.~5 in~\cite{jayaweera} varies roughly from around $1$ to $5$ bits/s/Hz as $N_R$ receive antennas increase from $4$ to $12$ respectively while the corresponding capacity values for these settings in~Fig.~\ref{fig:one} (or~Fig.~\ref{fig:three}) in this note range nearly from $1.18$ to $2.17$ bits/s/Hz respectively.
\item Though less noticeable, at high SNRs, the capacity values with CSIT in the~Fig.~5 in~\cite{jayaweera} are slightly lower in comparison to the corresponding values in~Fig.~\ref{fig:one} here.
\item Furthermore, the capacity curves \emph{without} CSIT in the~Fig.~5 and Fig.~6 in~\cite{jayaweera} are inaccurate; for example, the $C \approx \widehat{C}$ approximation at high SNR whenever $m = N_T$ (as noted in Eq.~\eqref{eq:cap:approxi}), is missing in the~Fig.~5 in~\cite{jayaweera}. Similar discrepancy can be noticed in the Fig.~6 in~\cite{jayaweera} by comparing capacity values of the $4 \times 4$ Rayleigh channel with and without CSIT at high SNRs (Eq.~\eqref{eq:cap:approxi} holds for this specific channel setting, but is missing in the Fig.~6 in~\cite{jayaweera} as well).
\end{itemize}
To summarize, there are serious inaccuracies in the ergodic capacity plots (both CSIT and no CSIT) presented in the~Fig.~5 and Fig.~6 in~\cite{jayaweera}, thus resulting in unwarranted comparison and conclusion. These plots are corrected in this note and presented as~Fig.~\ref{fig:one} and Fig.~\ref{fig:two} along with detailed justifications.

\begin{figure}[H]
\centering
\includegraphics[scale = 1.1]{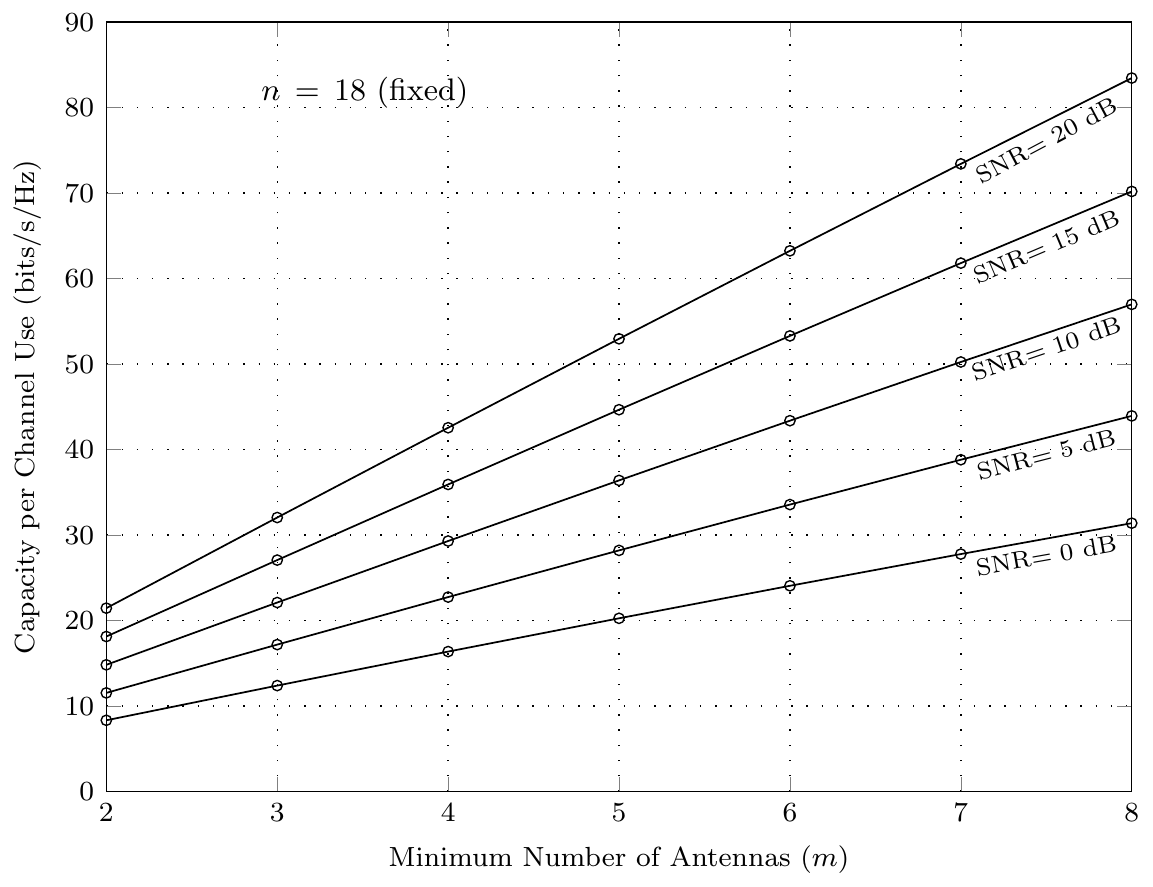}
\caption{Ergodic capacity versus minimum number of antennas $(m)$ with $n = 18$ (fixed).}
\label{fig:five}
\end{figure}

Fig.~\ref{fig:five} in this note is a corrected version of the Fig.~8 in~\cite{jayaweera}. On comparing the capacity values in these figures, the discrepancies can be noticed at large `minimum number of antennas $m$' for high SNR values.
It is difficult to identify all the sources of numerical errors in the ergodic capacity plots in~\cite{jayaweera}. One error source is the discrepancy in the threshold value computed in~\cite{jayaweera}. Eq.~(36) in~\cite{jayaweera} is reproduced below:
\begin{align}\label{eq:cutoff1}
\int_{\gamma_0}^{\infty}\left( \dfrac{1}{\gamma_0} -\dfrac{1}{\gamma}\right) f_{\boldsymbol{\gamma}}(\gamma) d \gamma = 1.
\end{align}
\begin{figure}[http]
\centering
\includegraphics[scale = 1.1]{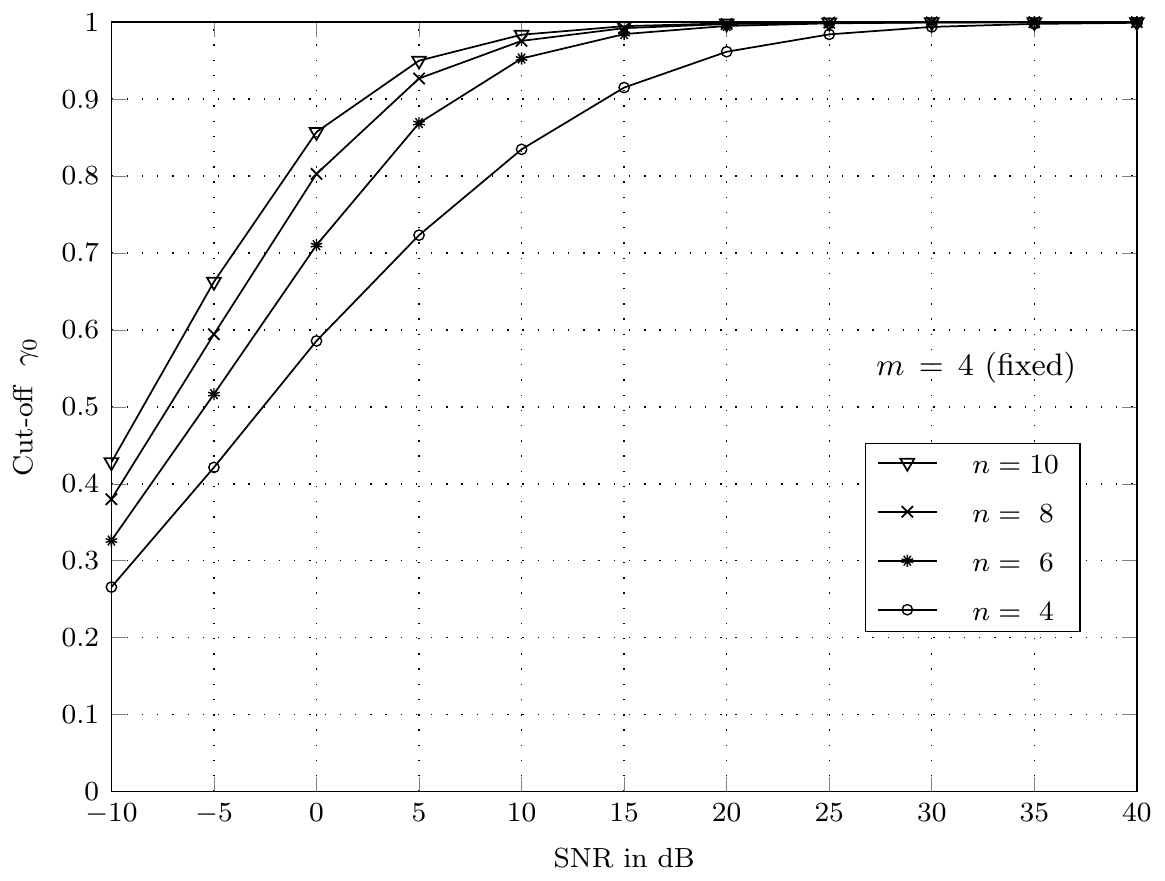}
\caption{Optimal cut-off versus SNR for $m = 4$.}
\label{fig:four}
\end{figure}
\noindent
Eq.~\eqref{eq:cutoff1} is a modification of Eq.~\eqref{eq:cutoff} with the substitution $\gamma = \lambda \tfrac{P}{mN_0}$ mentioned in the third
footnote. In~\cite{jayaweera}, this cut-off $\gamma_0$ is computed and plotted in the Fig.~4 for $m = 4$ (fixed), $n = 4,\,6,\,8,\,10$ and SNR (in dB scale) $= -10,\,-5,\,0,\,5,\,10,\dots,\,40$.
For these exact channel settings, we compute the cut-off $\gamma_0$ and plotted the values in Fig.~\ref{fig:four} in this note. On comparing these two figures, we conclude that the cut-off values for the selected SNR range in~\cite{jayaweera} are lower than the corresponding cut-off values in Fig.~\ref{fig:four} in this note. The error gap (or discrepancy) is significantly larger at low SNR and narrows down at high SNR.
\begin{remark}
In~\cite{jayaweera}, Section~III is devoted to ergodic capacity evaluation for the single receiver antenna systems i.e. $N_R = 1$. The $\gamma_0$ plots (Fig.~1) and the capacity curves (both CSIT and no CSIT) (Fig.~2) in~\cite{jayaweera} for the rank-one channel are `correct'. Also, the ergodic rates in the  Fig.~3 in~\cite{jayaweera} are correct.
\end{remark}

\begin{figure}[http]
\centering
\includegraphics[scale = 1.1]{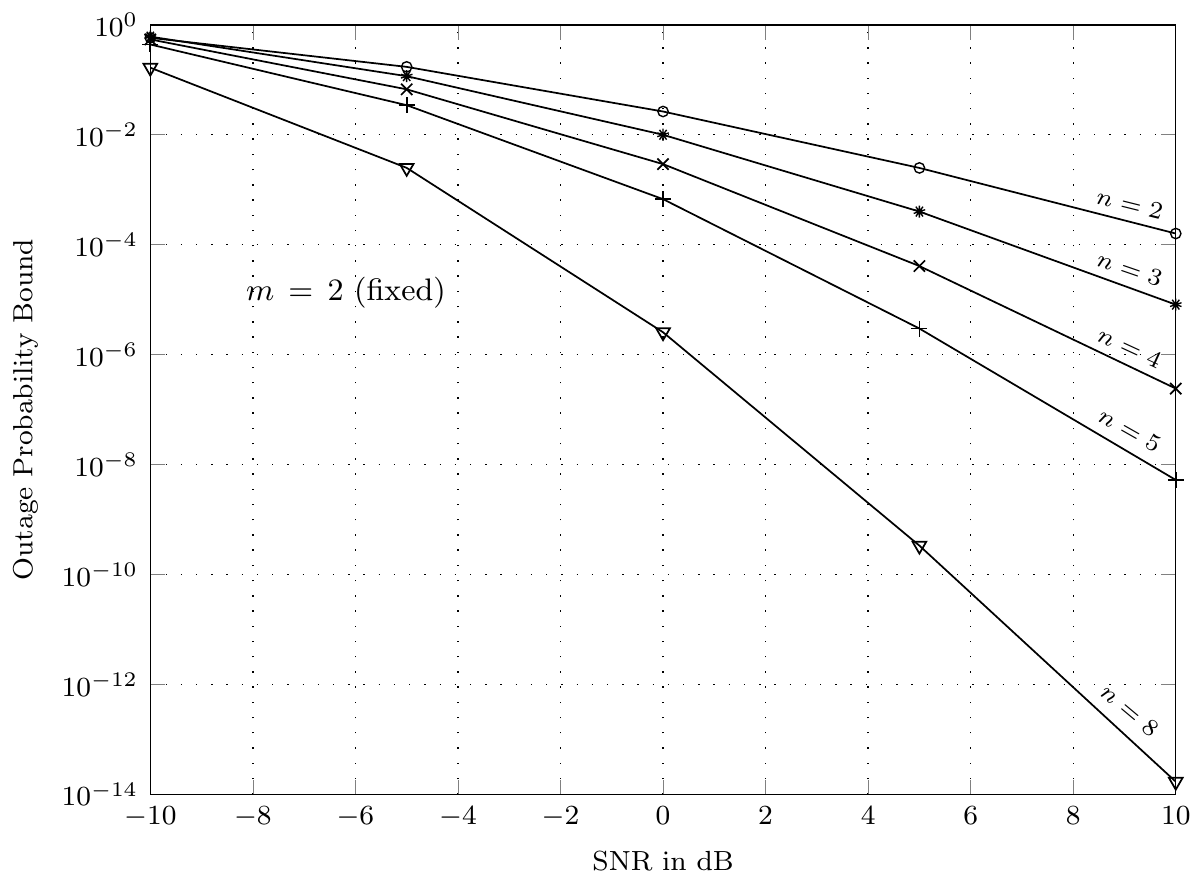}
\caption{Upper bound (Eq.~\eqref{eq:bound1}) for outage probability of a coherent MIMO IID Rayleigh channel with CSIT and $m = 2$ (fixed). This figure is a corrected version of the Fig.~9 in~\cite{jayaweera}.}
\label{fig:six}
\end{figure}

\begin{figure}[http]
\centering
\includegraphics[scale = 1.1]{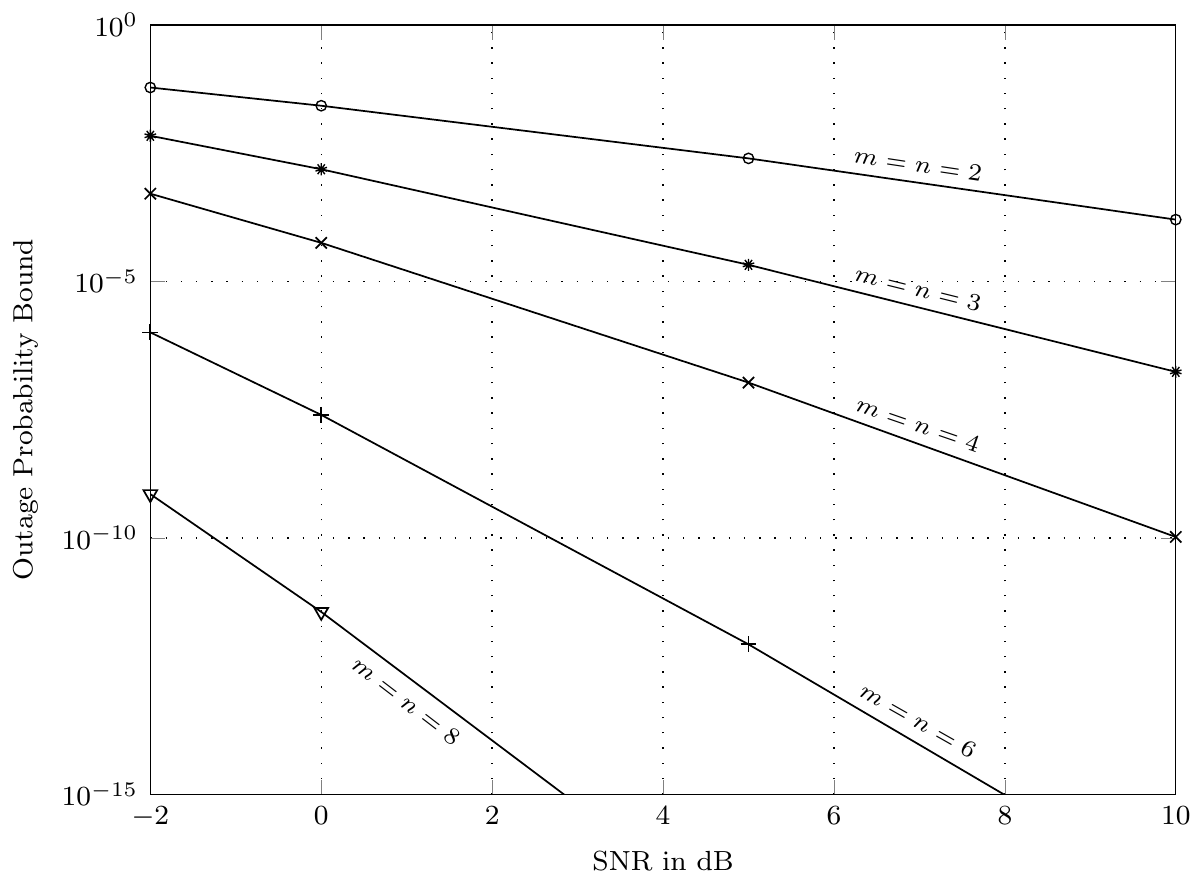}
\caption{Upper bound (Eq.~\eqref{eq:bound2}) for outage probability of a coherent MIMO IID Rayleigh channel with CSIT and $m = n$. This figure is a corrected version of the Fig.~10 in~\cite{jayaweera}.}
\label{fig:seven}
\end{figure}

\begin{figure}[http]
\centering
\includegraphics[scale = 1.1]{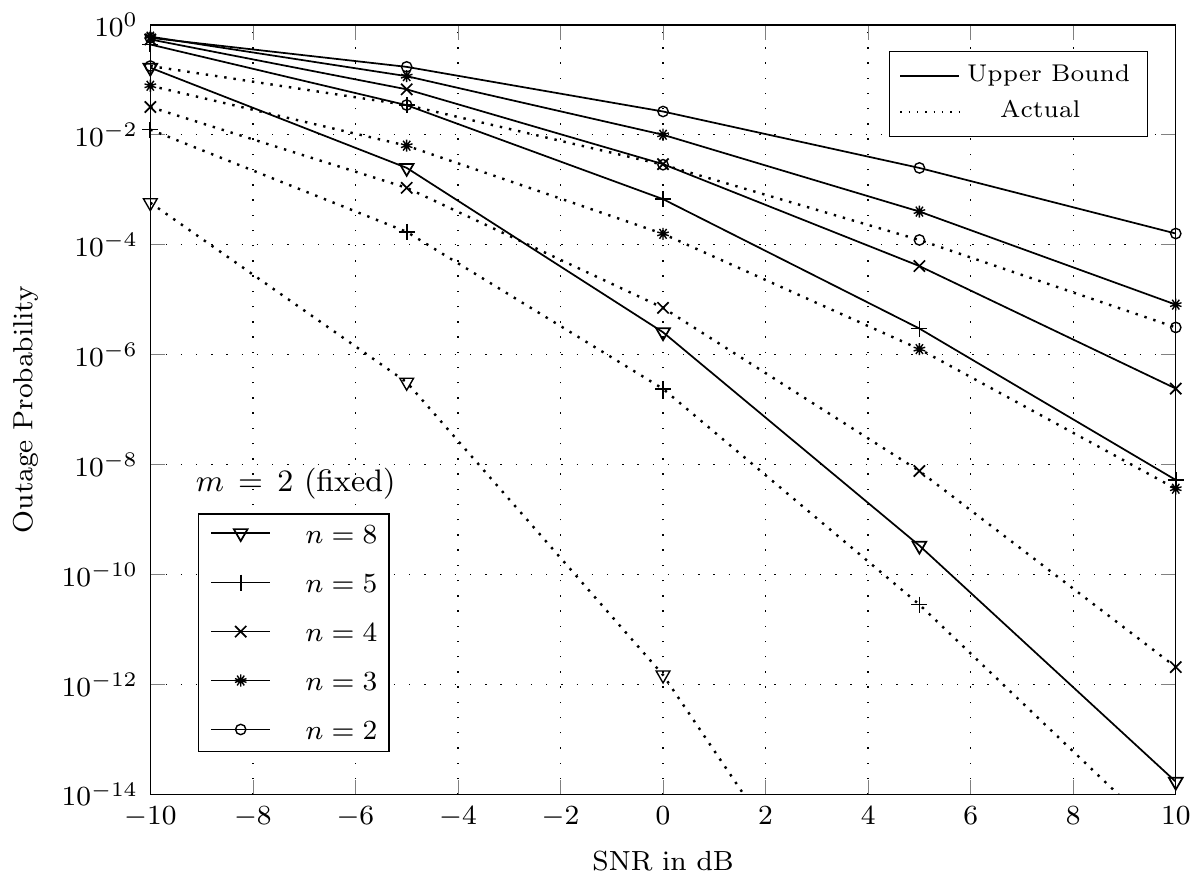}
\caption{Upper bound (Eq.~\eqref{eq:bound1}) compared with actual outage probability of a coherent MIMO IID Rayleigh channel with CSIT and $m = 2$ (fixed).}
\label{fig:eight}
\end{figure}

\begin{figure}[http]
\centering
\includegraphics[scale = 1.1]{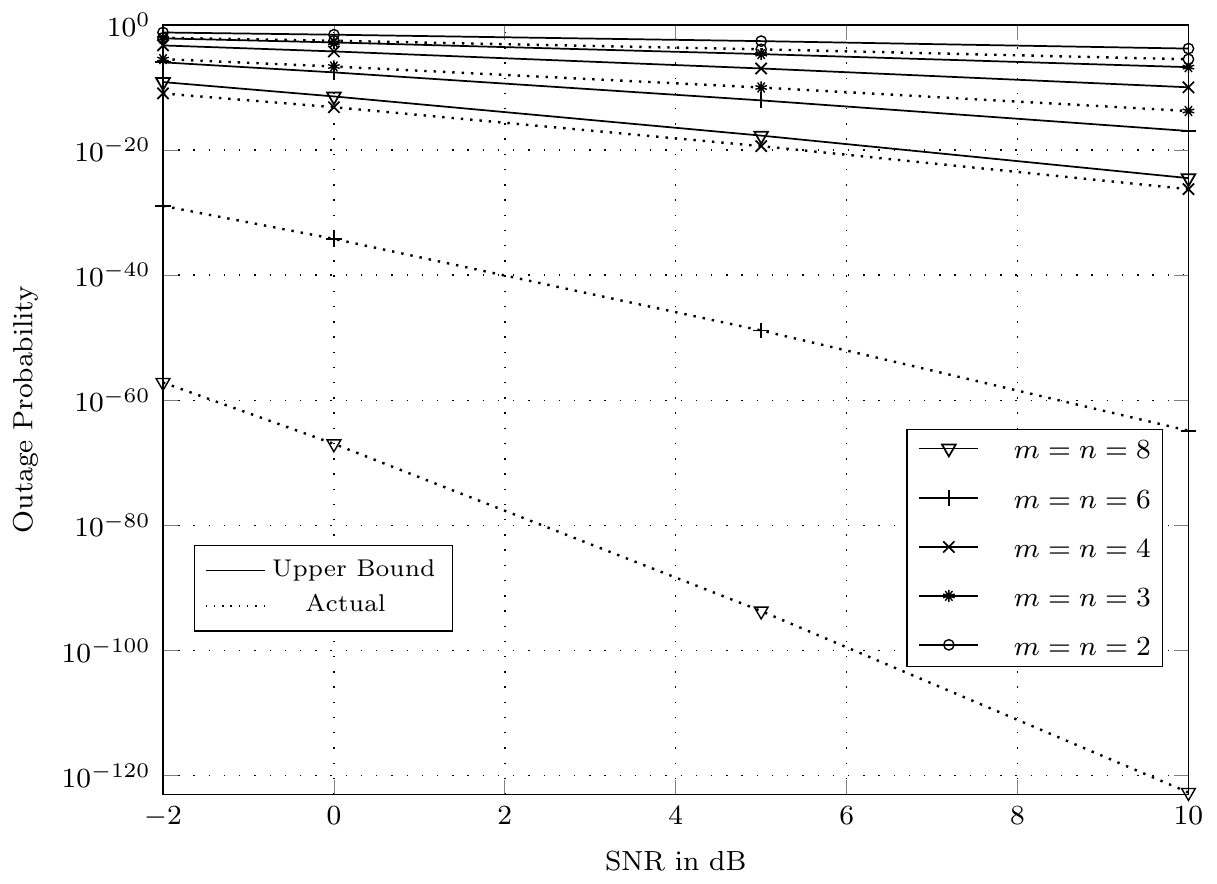}
\caption{Upper bound (Eq.~\eqref{eq:bound2}) compared with actual outage probability of a coherent MIMO IID Rayleigh channel with CSIT and $m = n$.}
\label{fig:nine}
\end{figure}

In~\cite{jayaweera}, Fig.~9 and Fig.~10 are plots of the proposed upper bounds (Eq.~(64) and Eq.~(66)) on the outage probability. The outage probability is given by $P_{\mathrm{out}}^{n,m} := F_{\boldsymbol{\lambda_{max}}} (\lambda_0)$ where $F_{\boldsymbol{\lambda_{max}}} (\cdot)$ denotes the CDF of the  strongest eigenmode $\boldsymbol{\lambda_{max}}$. For the ease of reading, we reproduce these bounds:
\begin{align}
P_{\mathrm{out}}^{n,m} \, &\leq \, \dfrac{1}{\Gamma(n)\Gamma(m)} [\Gamma(n+m-1) - \Gamma(n+m-1,\lambda_0)] \equiv p_1,\label{eq:bound1}\\
P_{\mathrm{out}}^{m,m} \, &\leq \, \min\{p_1,\,p_2\},\label{eq:bound2}
\end{align}
where $p_2 \equiv 1 - e^{-m\lambda_0}$. These bounds are evaluated incorrectly in~\cite{jayaweera} due to errors in the cut-off values; the  cut-off $\gamma_0$ and the waterfilling threshold $\lambda_0$ are related as $\gamma_0 = \tfrac{P}{m}\lambda_0$ (Eq.~(17) in~\cite{jayaweera}). Fig.~\ref{fig:six} and Fig.~\ref{fig:seven} presented in this note are corrected versions of the Fig.~9 and Fig.~10 respectively in~\cite{jayaweera}. Rather, the outage probability can be computed directly using the $F_{\boldsymbol{\lambda_{max}}} (\cdot)$ distribution function~\cite[Corollary~2]{kang}. For the sake of completeness, we compare the upper bounds in Eq.~\eqref{eq:bound1} and Eq.~\eqref{eq:bound2} with the actual outage probability in Fig.~\ref{fig:eight} and Fig.~\ref{fig:nine} respectively in this note.


\newpage
\section{Discussion}
An interesting contrast between performances of the coherent MIMO systems with and without CSIT can be inferred from the corrected capacity curves in Fig.~\ref{fig:one} and Fig.~\ref{fig:two}:
\begin{itemize}
\item For the case when the number of receive antennas $N_R$ is kept fixed and number of transmit antennas $N_T$ is increasing (above $N_R$), the capacity curves without CSIT suggest the effect of varying $N_T$ is marginal. To justify this, notice that the transmit power is spread out equally across all directions in the $\mathcal{C}^{N_T}$ vector space due to the lack of CSIT, leading to `wasted' energy which increases with increasing $N_T$. On the contrary, the diversity gains possible to the receiver improve due to increasing $N_T$. Overall, the improvement in the capacity without CSIT is marginal; with CSIT, the vector is projected on only the desirable $m$ non-zero eigenvalues or directions.
\item In contrast, for the case when $N_T$ is fixed and $N_R$ is increasing (above $N_T$), the gap between the capacity curves with CSIT and without CSIT is generally small and nearly vanishes at high SNR~(see Eq.~\eqref{eq:cap:approxi}). At high SNRs, and since $m = N_T$ holds here, it is easy to deduce that the optimal power allocations at the transmit antennas in both situations (CSIT and no CSIT) are nearly identical. Furthermore, with focus on the capacity without CSIT here, notice that the spread of transmit power or energy across all directions in $\mathcal{C}^{N_T}$ remains `same' (since $N_T$ is fixed) while the \emph{coherent} receiver is able to exploit higher diversity gains possible due to increasing $N_R$.
\end{itemize}
This overall contrast is somewhat limited in~\cite{jayaweera} because of flaws in the capacity curves.

The implications of these observations for the coherent MIMO system design can be briefly summarized as follows:
\begin{enumerate}
\item the loss of capacity due to lack of channel state information at the transmitter side for the MIMO Rayleigh fading channel can be made significantly smaller at \emph{medium to high} SNRs by providing a `larger' antenna array at the receiver\footnote{number of receive antennas to be larger than that of transmit antennas}.
\item On the other hand, larger antenna array at the transmitter yields minimal improvement in the capacity for the lack of CSIT. Alternatively, when the antenna array at the transmitter side is larger than at the receiver end, it is imperative to provide channel
    information at the transmitter to increase the throughput significantly.
\end{enumerate}


%

\balance
\begin{appendices}
\section{Ergodic capacity (Without CSIT)}\label{sec:app:A}
\begin{table}[H]
\caption{$N_R = 4$, $N_T = 4$ }\label{tab:fading6}
\centering
\begin{tabular}{c | c | c}
\hline 
SNR  &  Capacity (Monte Carlo) &  Capacity (Numerical) \\
(dB)   &(bits/s/Hz) & (bits/s/Hz) \\[.50ex]
\hline
-15&0.653249&0.653210\\\hline
-10&1.731278&1.731566\\\hline
-5&3.901254&3.901646\\\hline
0&7.360472&7.360570\\\hline
5&11.954841&11.958356\\\hline
10&17.402147&17.398369\\\hline
\end{tabular}
\end{table}
\begin{table}[H]
\caption{$N_R = 4$, $N_T = 6$ }\label{tab:fading7}
\centering
\begin{tabular}{c | c | c}
\hline 
SNR  &  Capacity (Monte Carlo) &  Capacity (Numerical) \\
(dB)  &(bits/s/Hz) & (bits/s/Hz) \\[.50ex]
\hline
-15&0.664150&0.664113\\\hline
-10&1.796232&1.796320\\\hline
-5&4.152699&4.153194\\\hline
0&8.004090&8.003924\\\hline
5&13.143525&13.143428\\\hline
10&19.114700&19.114446\\\hline
\end{tabular}
\end{table}
\begin{table}[H]
\caption{$N_R = 4$, $N_T = 8$ }\label{tab:fading8}
\centering
\begin{tabular}{c | c | c}
\hline 
SNR   &  Capacity (Monte Carlo) &  Capacity (Numerical) \\
(dB) &(bits/s/Hz) & (bits/s/Hz) \\[.50ex]
\hline
-15&0.669718&0.669736\\\hline
-10&1.830347&1.830585\\\hline
-5&4.287120&4.287054\\\hline
0&8.331369&8.330539\\\hline
5&13.683868&13.685661\\\hline
10&19.806269&19.806603\\\hline
\end{tabular}
\end{table}
\begin{table}[H]
\caption{$N_R = 4$, $N_T = 10$ }\label{tab:fading9}
\centering
\begin{tabular}{c | c | c}
\hline 
SNR  &  Capacity (Monte Carlo) &  Capacity (Numerical) \\
(dB)  &(bits/s/Hz) & (bits/s/Hz) \\[0.50ex]
\hline
-15&0.673125&0.673167\\\hline
-10&1.851860&1.851785\\\hline
-5&4.369483&4.369846\\\hline
0&8.525321&8.525929\\\hline
5&13.991696&13.992216\\\hline
10&20.179583&20.179123\\\hline
\end{tabular}
\end{table}
\begin{table}[H]
\caption{$N_R = 4$, $N_T = 12$ }\label{tab:fading10}
\centering
\begin{tabular}{c | c | c}
\hline 
SNR  &  Capacity (Monte Carlo) &  Capacity (Numerical) \\
(dB)  &(bits/s/Hz) & (bits/s/Hz) \\[.50ex]
\hline
-15&0.675493&0.675479\\\hline
-10&1.866260&1.866194\\\hline
-5&4.426025&4.426032\\\hline
0&8.655543&8.655479\\\hline
5&14.188157&14.188736\\\hline
10&20.413271&20.412153\\\hline
\end{tabular}
\end{table}
\begin{table}[H]
\caption{$N_R = 6$, $N_T = 4$ }\label{tab:fading11}
\centering
\begin{tabular}{c | c | c}
\hline 
SNR   &  Capacity (Monte Carlo) &  Capacity (Numerical) \\
(dB) &(bits/s/Hz) & (bits/s/Hz) \\[.50ex]
\hline
-15&0.956394&0.956307\\\hline
-10&2.462461&2.462844\\\hline
-5&5.339700&5.340373\\\hline
0&9.687900&9.688512\\\hline
5&15.175786&15.175911\\\hline
10&21.334315&21.332693\\\hline
\end{tabular}
\end{table}
\begin{table}[H]
\caption{$N_R = 8$, $N_T = 4$ }\label{tab:fading12}
\centering
\begin{tabular}{c | c | c}
\hline 
SNR  &  Capacity (Monte Carlo) &  Capacity (Numerical) \\
(dB)  &(bits/s/Hz) & (bits/s/Hz) \\[.50ex]
\hline
-15&1.245405&1.245477\\\hline
-10&3.122625&3.123057\\\hline
-5&6.537848&6.537777\\\hline
0&11.433137&11.432010\\\hline
5&17.310972&17.311652\\\hline
10&23.673682&23.672450\\\hline
\end{tabular}
\end{table}
\begin{table}[H]
\caption{$N_R = 10$, $N_T = 4$ }\label{tab:fading13}
\centering
\begin{tabular}{c | c | c}
\hline 
SNR   &  Capacity (Monte Carlo) &  Capacity (Numerical) \\
(dB) &(bits/s/Hz) & (bits/s/Hz) \\[0.50ex]
\hline
-15&1.521897&1.521850\\\hline
-10&3.722394&3.723170\\\hline
-5&7.553039&7.552851\\\hline
0&12.800568&12.799501\\\hline
5&18.881360&18.882073\\\hline
10&25.330437&25.329254\\\hline
\end{tabular}
\end{table}
\begin{table}[H]
\caption{$N_R = 12$, $N_T = 4$ }\label{tab:fading14}
\centering
\begin{tabular}{c | c | c}
\hline 
SNR  &  Capacity (Monte Carlo) &  Capacity (Numerical) \\
(dB)  &(bits/s/Hz) & (bits/s/Hz) \\[.50ex]
\hline
-15&1.786068&1.786435\\\hline
-10&4.272003&4.272138\\\hline
-5&8.429043&8.428782\\\hline
0&13.916139&13.916232\\\hline
5&20.120091&20.119217\\\hline
10&26.612370&26.612935\\\hline
\end{tabular}
\end{table}

\section{Ergodic rates with CSIT (On-off \& Waterfilling Power scheme)}\label{sec:app:B}

\begin{table}[H]
\caption{$N_R = 4$, $N_T = 4$}\label{tab:fading15}
\centering
\begin{tabular}{c | c | c | c}
\hline 
SNR  &  Capacity (Waterfilling) &  On-off (Monte-carlo) & On-off (numerical) \\
(dB) &(bits/s/Hz) & (bits/s/Hz)  & (bits/s/Hz) \\[.50ex]
\hline
0&8.141455&5.270533&5.264009331\\\hline
-5&4.788881&3.691180&3.684889873\\\hline
-10&2.519896&2.259619&2.254272476\\\hline
-15&1.182636&1.145351&1.141753736\\\hline
-20&0.500745&0.485424&0.483608496\\\hline
-25&0.195009&0.187945&0.187270682\\\hline
-30&0.072300&0.069711&0.069370405\\\hline
-35&0.025994&0.025088&0.024979542\\\hline
-40&0.009099&0.008847&0.008820817\\\hline
\end{tabular}
\end{table}

\begin{table}[H]
\caption{$N_R = 4$, $N_T = 6$}\label{tab:fading16}
\centering
\begin{tabular}{c | c | c | c}
\hline 
SNR  &  Capacity (Waterfilling) &  On-off (Monte-carlo) & On-off (numerical) \\
(dB) &(bits/s/Hz) & (bits/s/Hz)  & (bits/s/Hz) \\[.50ex]
\hline
0&9.974868&5.683648&5.677020344\\\hline
-5&5.894041&4.083595&4.077349245\\\hline
-10&3.107786&2.599575&2.593966946\\\hline
-15&1.461355&1.384511&1.380272409\\\hline
-20&0.619716&0.603249&0.600875364\\\hline
-25&0.241313&0.233272&0.232197311\\\hline
-30&0.088662&0.085671&0.085386323\\\hline
-35&0.031695&0.030700&0.030499115\\\hline
-40&0.011062&0.010743&0.010684742\\\hline
\end{tabular}
\end{table}

\begin{table}[H]
\caption{$N_R = 4$, $N_T = 8$}\label{tab:fading17}
\centering
\begin{tabular}{c | c | c | c}
\hline 
SNR  &  Capacity (Waterfilling) &  On-off (Monte-carlo) & On-off (numerical) \\
(dB) &(bits/s/Hz) & (bits/s/Hz)  & (bits/s/Hz) \\[.50ex]
\hline
0&11.527083&5.985882&5.979340923\\\hline
-5&6.859201&4.374472&4.368011005\\\hline
-10&3.633938&2.859162&2.853247814\\\hline
-15&1.714000&1.578669&1.574165663\\\hline
-20&0.727666&0.708751&0.706173821\\\hline
-25&0.283272&0.274775&0.273769589\\\hline
-30&0.103875&0.100618&0.100242818\\\hline
-35&0.036873&0.035746&0.035616903\\\hline
-40&0.012848&0.012466&0.012410698\\\hline
\end{tabular}
\end{table}

\begin{table}[H]
\caption{$N_R = 4$, $N_T = 10$}\label{tab:fading18}
\centering
\begin{tabular}{c | c | c | c}
\hline 
SNR  &  Capacity (Waterfilling) &  On-off (Monte-carlo) & On-off (numerical) \\
(dB) &(bits/s/Hz) & (bits/s/Hz)  & (bits/s/Hz) \\[.50ex]
\hline
0&12.836196&6.226041&6.219488459\\\hline
-5&7.731536&4.606987&4.600511855\\\hline
-10&4.119671&3.070413&3.064576730\\\hline
-15&1.949462&1.743438&1.738625987\\\hline
-20&0.829473&0.805049&0.802063549\\\hline
-25&0.323232&0.314441&0.313158002\\\hline
-30&0.118232&0.114694&0.114363686\\\hline
-35&0.041907&0.040672&0.040480673\\\hline
-40&0.014515&0.014124&0.014049351\\\hline
\end{tabular}
\end{table}

\begin{table}[H]
\caption{$N_R = 4$, $N_T = 12$}\label{tab:fading19}
\centering
\begin{tabular}{c | c | c | c}
\hline 
SNR  &  Capacity (Waterfilling) &  On-off (Monte-carlo) & On-off (numerical) \\
(dB) &(bits/s/Hz) & (bits/s/Hz)  & (bits/s/Hz) \\[.50ex]
\hline
0&13.933972&6.426369&6.419435968\\\hline
-5&8.531219&4.801641&4.795005000\\\hline
-10&4.574472&3.249739&3.243676350\\\hline
-15&2.172983&1.887175&1.882104161\\\hline
-20&0.925992&0.893189&0.890072923\\\hline
-25&0.361619&0.352384&0.350926130\\\hline
-30&0.132287&0.128491&0.127962447\\\hline
-35&0.046683&0.045373&0.045165040\\\hline
-40&0.016133&0.015711&0.015626368\\\hline
\end{tabular}
\end{table}

\end{appendices}

\end{document}